\documentclass[11pt, oneside]{article}   	
\usepackage{geometry}                		
\usepackage{xcolor}
\usepackage{graphicx}				
\usepackage{amssymb}
\usepackage{url}
\usepackage{multicol}
\usepackage{cite}
\usepackage[unicode=true,colorlinks = true, linkcolor = blue, urlcolor = blue, citecolor = blue, anchorcolor = blue]{hyperref} 
\usepackage{siunitx}           

\title{Status of the International Linear Collider}
\author{ICFA ILC International Development Team and ILC-Japan\footnote{All the members are listed at the end of this document.}\\[5 mm]
Revised from the document submitted on 31 March 2025 \\
for the European Strategy for Particle Physics Update 2026}

\date{\small First revision on 16 May 2025 \\ Second revision on 26 May 2025 \\ Third revision on 25 May 2025\\(submitted on 26 May 2025 for EPPSU2026 as an update) \\Fourth revision on 5 June 2025}

\begin{document}
\maketitle
\vspace{3cm}
\begin{abstract}
\noindent 
This paper is not a proposal for a CERN future project but provides information on the International Linear Collider (ILC) considered for Japan in order to facilitate the European Strategy discussion in a global context. It describes progress to date, ongoing engineering studies, updated cost estimate for the machine at $\sqrt{s}=$\SI{250}{GeV} and the situation in Japan. The physics of the ILC is not presented here, but jointly for all Linear Collider projects in a separate document ``A Linear Collider Vision for the Future of Particle Physics'' submitted for the forthcoming European Strategy deliberations. 
\end{abstract}

\newpage
\pagenumbering{arabic}\setcounter{page}{1}

\section{Introduction}
The International Linear Collider, the ILC, is a proposed $e^+e^-$ linear collider based on well-established superconducting radiofrequency (SRF) technology with a mature technical design and offering polarized beams. It will be a capable electron-positron Higgs factory in its first phase, at $\sqrt{s}=$\SI{250}{GeV}, and could be upgraded to \SI{550}{GeV}, and
to 1\,TeV, using the same technology. The scientific programme offered by such a machine, along with various upgrade opportunities, is presented in~Ref.~\cite{LinearColliderVision:2025hlt}.

The ILC  has been developed as a global project, supported by the International Committee for Future Accelerators (ICFA) through a succession of Subpanels since the ILC's conception at the time of the merging of three linear collider efforts: TESLA by DESY in Germany, JLC in Japan and NLC by SLAC in the US. It was defined as an SRF technology based machine with a collision energy of $\sqrt{s}=$\SI{500}{GeV}. After the discovery of the Higgs boson in 2012~\cite{bib-Higgs}, the Japanese Association of High Energy Physicists (JAHEP) proposed~\cite{bib-JAHEP} to host the ILC in Japan as a global project, starting at $\sqrt{s}=$\SI{250}{GeV}. 

The ILC  Technical Design Report (TDR)~\cite{bib-TDR, Baer:2013cma, Adolphsen:2013jya, Adolphsen:2013kya, Behnke:2013lya}, completed in 2013, provided a detailed technical description of the machine. The SRF technology has been adopted by several accelerators, including the successfully operating European XFEL (Eu.XFEL) in Hamburg\footnote{\url{https://www.xfel.eu/index_ger.html}}, and the maturity of the technology has been well demonstrated. International efforts at the Accelerator Test Facility (ATF) at KEK\footnote{\url{https://atf.kek.jp/atfbin/view/Main/WebHome}} have demonstrated that a very strongly focused beam (nano-beam), which is one of the essential points for reaching high luminosities, can be achieved. 

ICFA set up the International Development Team (IDT)\footnote{\url{http://icfa.hep.net/wp-content/uploads/ICFA_IDT_Structure.pdf}, \url{https://linearcollider.org/}} in 2020 as a subpanel to prepare the engineering design  phase of the ILC. Hosted by KEK, the IDT has been collaborating with the Japanese and global ILC community to advance the realisation of the ILC. The IDT has also been playing a pivotal role in supporting the physics efforts by the worldwide linear collider community, since the last update of the European Strategy for Particle Physics. Its activities include providing software tutorials, organising physics discussions and mini-workshops on detector technologies, coordinating conference appearances, and facilitating the organisation of the annual Linear Collider Workshops. 

This document summarises the following aspects of the IDT's activities: the Pre-lab proposal and the work of the International Expert Panel (Section~2), the ILC Technology Network (Section~3), and the 2024 update of the cost estimate for the ILC in Japan (Section~4). Section~5 covers construction timelines, environmental considerations, and annual power consumption. Some of these topics also provide key input to the Linear Collider Facility proposal for CERN~\cite{LinearCollider:2025lya}. Section~6 provides a brief overview of the ILC's status in Japan, followed by the conclusion in Section~7.

\section{International Development Team}\label{section-2}
The IDT organized working groups on accelerator development and on physics and detectors and convened workshops. 
In 2021, the IDT published a proposal~\cite{bib-prelab} for the ILC Preparatory Laboratory (Pre-lab). 

The Pre-lab would be the base for completing the engineering design of the ILC. The proposal described the organisation, work plan and required resources. Reflecting the global nature of the ILC, the proposed Pre-lab organisation was distributed across laboratories worldwide with a small headquarters in Japan. It was suggested that some 
indication of interest from Japan to consider hosting the ILC would be needed to initiate the Pre-lab. 

The Pre-lab proposal was evaluated by the ad-hoc MEXT ILC Advisory Panel. The panel's conclusion in 2022~\cite{bib-MEXT} was that it would be premature to proceed with the Pre-lab, 
which required 
an expression of interest to host, while the panel's concerns about the ILC itself remain unresolved. 
The panel's concerns included the impact of ILC's high cost (compared to science projects in Japan to date), uncertainties in the level of foreign contributions and the lack of broad support across Japanese academia. Nonetheless, further development of advanced accelerator technology was encouraged, while keeping the site issue on hold. 

Following the guidance from the MEXT ILC Advisory Panel, the IDT took two actions. The first was to establish a framework to advance time-critical research and development (R\&D) work by leveraging broad interest of laboratories worldwide in ILC technology, thereby maintaining momentum toward the ILC project. This framework, known as the ILC Technology Network (ITN), already supports several ongoing activities, as described in the following section.

The second action was organising the International Expert Panel (IEP) to understand the root causes behind the lack of political progress
in eliciting a positive response from 
the Japanese government regarding its potential interest in hosting the ILC in Japan. The conclusion of the IEP is that the Japanese government and the governments of potential major partner nations differ in their views on the first step of the global decision-making process for constructing the ILC. 
According to the IEP analysis, the potential partner countries are of the opinion that the Japanese government should initiate a global discussion by first expressing an interest to host the ILC in Japan. This expression of interest would then be followed by intergovernmental discussions about participation.
However, this view is in contrast with the 
Japanese government's (i.e. MEXT's) expectation that 
an international agreement to construct the ILC, not tied to a specific site, should precede any decision about hosting. 

The IEP and IDT acknowledge the position taken by MEXT. 
Nonetheless, it is hoped that Japan would consider initiating international discussions towards achieving MEXT's envisaged first step - an international consensus to construct the ILC.

\section{ILC Technology Network}
\begin{figure}[htb]
\begin{center}
\includegraphics[width=0.8\textwidth]{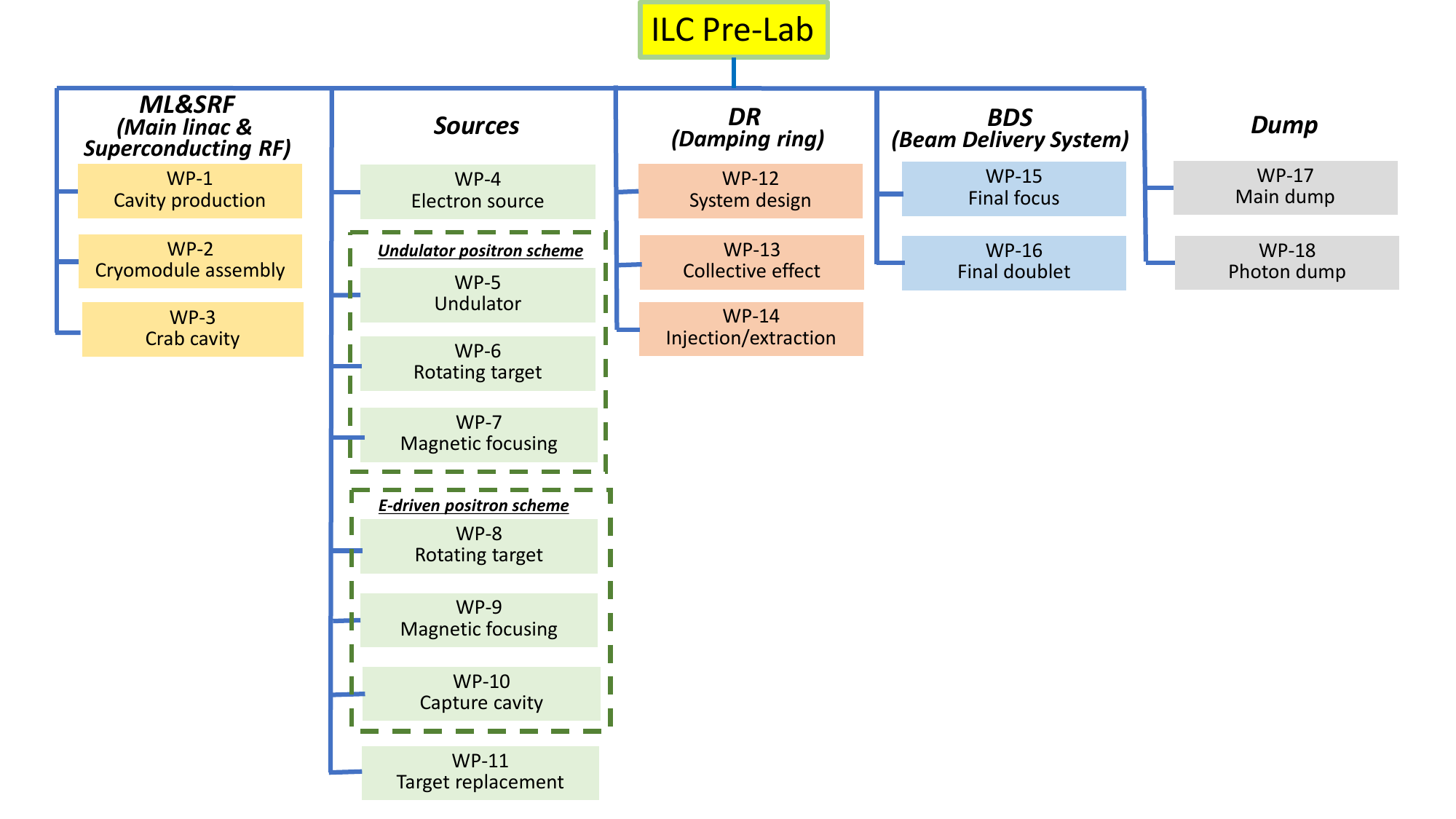}
\caption{Work Packages described in the Pre-lab proposal. The ITN selected 15 of them for the ITN work.}
\label{Prelab-WP}
\end{center}
\end{figure}

The Pre-lab proposal identified the remaining tasks needed for an engineering design for the ILC and compiled them as Work Packages (WPs) as shown in Figure~\ref{Prelab-WP}. 

Once it had become clear that the ILC Pre-lab could not be realised soon, the IDT identified particularly important and time-consuming tasks~\cite{ITN-WP} and started to address them, by setting up the ILC Technology Network (ITN),  
 in 2023. Those tasks include demonstrating the fabrication of SRF cavities in three regions of the world using state-of-the-art surface treatment methods, demonstrating techniques to achieve stable nano-beams over long periods of time and advancing the technical design study for two technologies of  positron production. 

Many of these issues are not only important for the ILC, but also relevant for a broad 
range
of applications with accelerators. They are often 
in
common with the subjects of interest 
in 
many laboratories. About 70 scientists and engineers from Japan, 10 from South Korea and 40 from Europe are already involved in the ITN work using the infrastructure of their laboratories.

The ITN guarantees the continuation of development work on SRF-based linear colliders, providing a smooth transition to the ILC Pre-lab or an analogous preparatory phase for any SRF-based linear collider in the future. It plays a key role in fostering international collaboration on linear colliders in the absence of an actual project and/or host laboratory.

In this section, the ITN work packages and their implementation status are described. 
\subsection{Work packages}
The ITN organised tasks as work packages in three areas, in a similar way to the Pre-lab proposal: superconducting radiofrequency acceleration (``SRF''), electron and positron sources (``Sources''), and ``Nano-beam''. The work packages were carefully selected from among the Pre-lab work packages and called Work Package-Prime (WPP)
to distinguish them from the WPs of the Pre-lab, although the WPPs carry the same identifying number as the Pre-lab WP. 
The WPPs are expected to be completed within two to four years from their start, i.e.\ between 2026 and 2028 for those that started right away when the ITN was established. Work packages start when interested laboratories commit to them. Pre-lab WP-13 on beam instabilities in the damping rings and WP-18 on the photon dump for the undulator-scheme positron source are not urgent, thus do not have corresponding ITN WPPs.

\paragraph{Superconducting Radiofrequency (SRF) }\hfill  \\
Three work packages are listed for the SRF area. 

\noindent {\bf WPP-1} 
is to demonstrate
that high-performance cavities can be manufactured worldwide. The ILC will require the manufacture of approximately 8000 SRF cavities, an order of magnitude more than existing accelerators, and the cavities are envisaged to be manufactured in three regions around the world. 

\noindent {\bf WPP-2} is related to cryomodule design. The design will be finalised for the vacuum-insulated vessel (cryomodule) that will house the SRF cavities to be operated at 2K and for peripheral equipment such as tuners.

\noindent {\bf WPP-3} concerns the crab cavity to be placed near the electron-positron collision point.
\paragraph{Electron and Positron Sources} \hfill \\
There is one work package for the electron source and six for the positron sources. 

\noindent {\bf WPP-4} relates to the electron source. The TDR adopts a high-voltage polarised electron source with a polarisation of 80\% or more. The key components of this electron source are the cold cathode, the laser to excite electrons and the high voltage insulation. Especially for the high-voltage insulation, the design will be further improved based on the experience of the past ten years.\\

Two schemes of positron sources, the 
undulator scheme and 
the 
electron-driven scheme, have been studied and developed. WP-5 in the Pre-lab proposal relates to the undulator itself; however, the undulator has been successfully prototyped and a large-scale undulator system has already been realised in the Eu.XFEL. Accordingly, WP-5 is not urgent and there is no corresponding ITN WPP-5.
 \\ 
\noindent {\bf WPP-6} concerns the target in undulator-scheme. The undulator-scheme positron source uses a \SI{230}{m}-long undulator to generate circularly polarised photons. The polarised photons are injected into a target to produce polarised positrons. To handle the heat load of the intense photon beam, a rotating target that is cooled by radiative cooling is used.

\noindent {\bf WPP-7} is related to the magnetic focusing system of the undulator scheme.  Polarised positrons are efficiently collected by a magnetic focusing system such as a pulsed solenoid. 

\noindent {\bf WPP-8} relates to the rotating target of the electron-driven scheme. 
In the electron-driven positron source, an electron beam from a \SI{3}{GeV} electron linear accelerator is injected into a rotating target to generate (unpolarized) positrons. 

\noindent {\bf WPP-9} concerns magnetic focusing in the electron-driven scheme. The positrons are focused by a flux concentrator. 

\noindent {\bf WPP-10} is related to the capture cavity of the electron-driven positron source.
Positrons are accelerated in the capture cavity. Especially in the acceleration cavity just after the magnetic focusing system, positrons in a mixture of electrons and positrons must be transported and accelerated as efficiently as possible. 
In addition, cooling must be considered because of the heat generated by primary particles from the target in the upstream area. 

\noindent {\bf WPP-11} is related to target replacement in both types of positron source. 
The target is scheduled to be replaced every few years due to deterioration. Due to high activation near the target, the target is being designed to be replaced remotely. 

\paragraph{Nano-beam}\hfill \\
Five work packages are listed for the nano-beam area, which includes the damping ring, the final focus system, and the beam dump.

\noindent {\bf WPP-12} is concerned with the system design of the damping ring. The ILC damping ring (DR) must simultaneously meet the requirements of low emittance and large dynamic aperture. By incorporating the latest light source design knowledge, the DR of the ILC can be further improved.

\noindent {\bf WPP-14} is related to the DR fast injection and extraction system. The beam extraction with a fast kicker system using semiconductor pulse power supplies with nano-second pulse length was demonstrated in principle at KEK's ATF about ten years ago. Since semiconductor technology continues to evolve, a fast kicker power supply using the latest semiconductor technology will be developed.

\noindent {\bf WPP-15} is related to the final focus. KEK's ATF has achieved 41 nm for the target beam size of 37 nm (equivalent to 7.7 nm at the ILC250). Further technological development, including long-time stability, is needed, and this will be addressed and promoted as a work package.

\noindent {\bf WPP-16} is related to vibration evaluation of superconducting magnets in the final focus area near the collision point.

\noindent {\bf WPP-17} is related to the main dump. A detailed design of the entire water dump system will be carried out.
\subsection{Implementation Status}
The ITN work packages are being shared among laboratories worldwide. In July 2023, KEK and CERN signed an agreement, stating that CERN will cooperate in ITN activities and serve as a hub for other European research laboratories. An agreement on the ITN has also been signed with Korea University in South Korea.

In October 2023, an ITN information meeting was held at CERN with around 70 participants from 28 research institutions in ten countries, including those attending remotely. In this meeting, participating laboratories identified their subjects of interest in the ITN work packages.
In July 2024, the second ITN information meeting was held in Tokyo, where the progress of each work package was reported. 

Regarding SRF, manufacturing of nine nine-cell cavities has started in Japan. Single-cell cavities are being manufactured in South Korea and Europe. A cryomodule including input couplers, tuners, magnetic shields, and quadrupole magnets will be designed by worldwide collaboration. One cryomodule will be manufactured in Japan, and some cavities, manufactured overseas as well as in Japan, will be installed into it. 

As for the positron source, work packages of the undulator-scheme have been started in Europe and of the electron-driven-scheme in Japan. 
Oxford University is responsible for DR injection and extraction work (WPP-14), which is synergistic with UK's Diamond Light Source upgrade. 
For the Final Focus System (WPP-15), the following three items are underway in parallel while some ATF equipment is being upgraded: 
wakefield mitigation, higher-order aberration correction, and training for ILC beam-tuning algorithms (with machine learning). 
Researchers from Europe with EAJADE\footnote{\url{https://www.eajade.eu/}} and from South Korea have been participating in ATF experiments since 2023.
Design study for the beam dump (WPP-17) is in progress in Japan.

 U.S. participation is planned to be under the framework of the existing U.S.-Japan Collaboration in HEP and is expected to be defined in the coming years. In the meantime, the U.S. laboratories interested in ITN activities will closely follow and participate in the discussion.
Additionally, the U.S. Department of Energy and National Science Foundation established a nationally coordinated U.S. Higgs Factory Coordination Consortium to advance the development of the U.S. program on accelerators and detectors for a future off-shore Higgs factory, following the recommendation of 
the 
Particle Physics Project Prioritization Panel (P5)\footnote{\url{https://www.usparticlephysics.org/2023-p5-report/}}.

\section{ILC cost update}
The cost of the ILC for a centre-of-mass energy of \SI{500}{GeV} (ILC500) was estimated at the time of the TDR in 2013. In 2017, the ILC cost was re-evaluated for an energy of \SI{250}{GeV}~\cite{bib-2017}, ILC250. Since then,
prices and currency exchange rates have evolved worldwide. For this reason, a new cost estimate has been made for ILC250, in view of the forthcoming update of the European Strategy for Particle Physics. The main specifications shown in Table~\ref{Specification} are identical to those in the 2017 cost re-evaluation \cite{bib-2017}. 
\begin{table}[htp]
\small
\caption{Main specification of the \SI{250}{GeV} ILC machine}
\begin{center}
\begin{tabular}{|c|c|c|c|c|c|c|}
\hline
$ L_{\rm tunnel}$ & ${\cal L }  $ & $\sqrt{s}$ &  $P_{\rm total}$ &  $f_{\rm collision}$ & $N_{\rm cavities}$& gradient\\
\hline
20.5 km & $1.35 \times 10^{34}~{\rm cm^{-2}s^{-1}}$ &250 GeV &111 MW & 5 Hz & $\sim 8000$ & 31.5 $\rm MV/m$\\ 
\hline
\end{tabular}
\end{center}
\label{Specification}
\end{table}%

The recently updated
estimate includes: (1)  the cost of the fabrication of accelerator components (Acc.) based on  Superconducting Radio Frequency (SRF) and standard technologies; (2) utilities and conventional facilities (CF); and (3) civil engineering (CE). The costs for preparatory work such as engineering design, for land acquisition and for local infrastructure such as roads, electricity, and water are not included. Cost of the experiments is also not included. Human resources, which includes laboratory staff and installation staff or workers, is unchanged from the estimate given in 2017. Only a high-level summary is presented here and comprehensive details are available in the backup document submitted alongside this report \cite{ILC-ESPP-ststus} for the ESPP 2026 update.
 
\subsection{Methodology and Procedure}\label{method}
Extensive efforts were made to collect up-to-date cost estimates (as of 2024) for SRF
by communicating with worldwide industrial partners and institutes. 
Price quotations received from multiple companies and institutes were averaged, rather than taking the lowest bid as done for the ILC500. 

For the CE estimate, the design and cost update followed the guidelines of the Japanese Government and the national tunnel costing standards, which have been updated annually by the Ministry of Land, Infrastructure and Transport\footnote{\url{https://www.mlit.go.jp/sogoseisaku/1_6_hf_000077.html}}. This costing method is a strict standard for tunnel construction in Japan. It has been well established and regarded as a reliable approach. Once a design is fixed, the cost is well determined by a dedicated bottom-up method. 

The new cost estimation for the remaining items was based on the ILC500 cost by applying appropriate scaling factors taking into account the price increases in the countries where the price quotations had been made. It should be noted that the ILC500 cost estimate was made based on the TDR design with a bottom-up approach.

For the 2024 cost estimate, the ILC currency unit was defined to be 1 ILC unit (ILCU) = 1 USD  as of January 2024. The ILC being a global project, contributions from partner countries are expected to be largely in-kind. Therefore, rather than using currency exchange rates, conversions from the estimates obtained in non-USD currencies to ILCU were based on Purchasing-Power-Parity (PPP) indices, as had been done for the ILC500 cost. PPP indices published by the Organisation for Economic Co-operation and Development (OECD)\footnote{\url{https://www.oecd.org/en/data/indicators/purchasing-power-parities-ppp.html}} as of January 2024, were used for Germany, Japan, and Switzerland. For the estimates from China, PPP indices published by the World Bank\footnote{\url{https://data.worldbank.org/}} were applied. The PPP conversion indices used in this cost update are summarised in Table \ref{PPP-table}. Machinery and Equipment (M\&E) index was used for the fabrication of the accelerator and conventional facility items. SRF Material index is used for superconducting material to be used to fabricate the superconducting cavities. 
\begin{table}[tp]
\caption{PPP conversion indices used in the cost in 2024.}
\begin{center}
\begin{tabular}{|l||c|c|c|c|c|}
\hline
Currency & USD & EUR & JPY & CHF & CNY \\
\hline \hline
M\&E & 1.0 & 0.91 & 129.6 & 1.05 & 8.83\\
\hline
SRF~Material & 1.0 & n/a & 149.1 & n/a & 7.1 \\
\hline
Source & OECD & OECD & OECD & OECD & W. Bank \\
\hline
\end{tabular}
\end{center}
\label{PPP-table}
\end{table}%

The cost estimate for civil engineering is kept in JPY, because this cost is generally assumed to be covered by Japan as the potential host country, and is very site-dependent.  
\subsection{Results}
\begin{figure}[!htb]
\begin{center}
\includegraphics[width=0.6\textwidth]{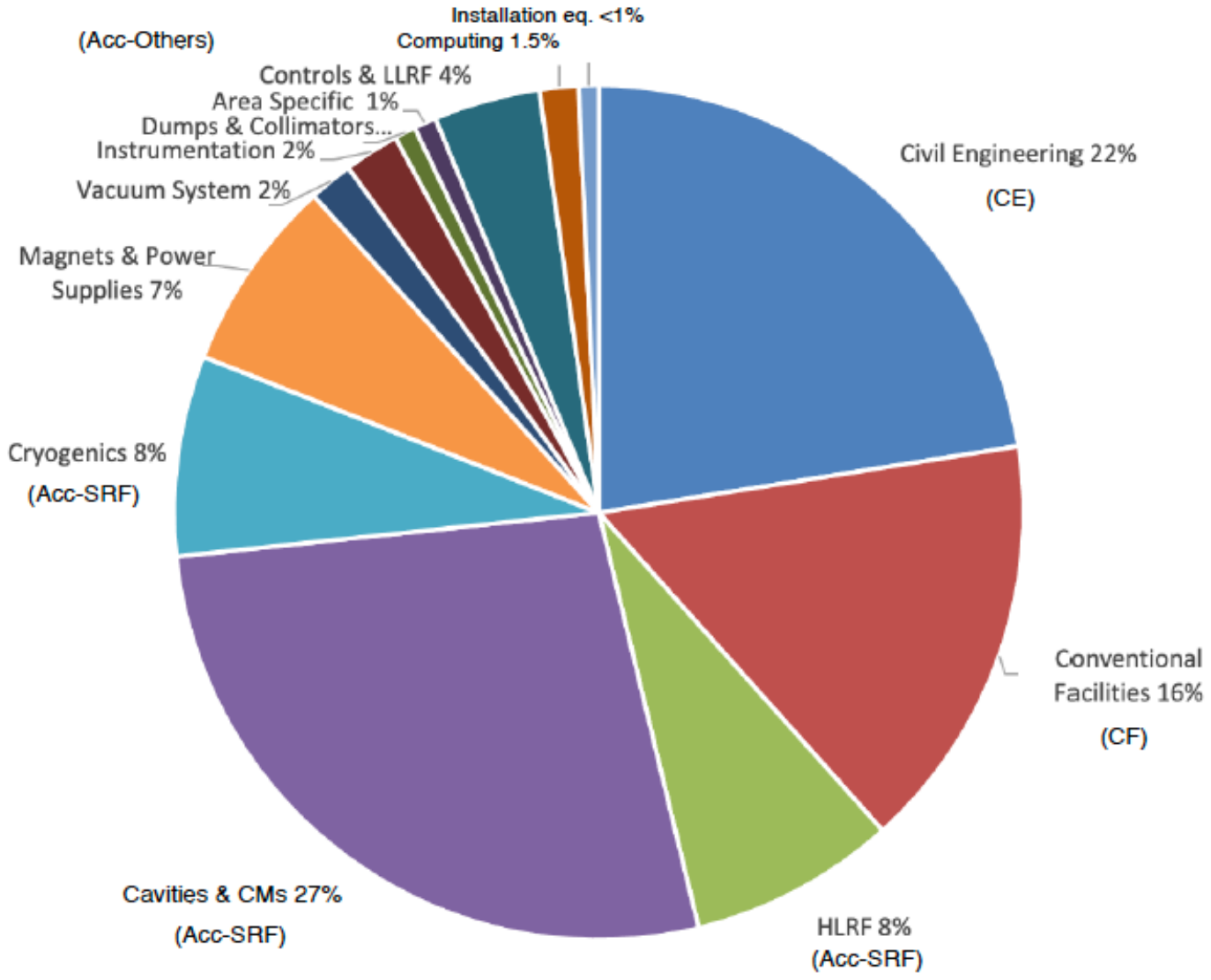}
\caption{Cost breakdown of the ILC 250. }
\label{cost=breakdown}
\end{center}
\end{figure}

The 2024 updated cost for ILC250 together with the required human resources are summarised in Table \ref{cost-table}. Figure~\ref{cost=breakdown} presents a more detailed breakdown. The cost of items corresponding to more than 70\% of the 2017 estimate for the \SI{250}{GeV} machine is now based on 2024 pricing, while the rest is based on the scaling described in Section~\ref{method}. Compared to the 2017 estimate, the cost for accelerator construction and conventional facilities has increased by $\sim 60 \%$, where $\sim 35\%$ is due to worldwide inflation and the rest reflects the change in production scaling, averaging multiple vendor cost-estimates, exchange rate variations, design updates, and other smaller changes. The cost for civil engineering work has increased by $\sim 50 \%$, where $\sim 30 \%$ is due to cost increases in Japan and the remainder is due to recent, more detailed  design work. Cost-estimate uncertainties are evaluated as $\sim 30 \%$. 

This result was reviewed by an international committee\footnote{  R. Brinkmann, J. Gao,  N.Holtkamp (Consultant), Ph. Lebrun, T. Raubenheimar, L. Rivkin (Chair)} with the conclusion that the estimated costs and uncertainties are reasonable. 
\begin{table}[ht]
\caption{The 2024 resource estimates for ILC construction}
\begin{center}
\begin{tabular}{|l||c|c|c|c|c|c|}
\hline
Items & \multicolumn{2}{c|}{Accelerator} & CF & CE &\multicolumn{2}{c|}{Human Resources}\\ \cline{2-3} \cline{6-7}

                & SRF & Others & & & @laboratories & installation \\
\hline
Estimate & 3.69 &  1.71  & 1.38 & 196  & 7.47     &  2.65   \\  \cline{2-7}
                &\multicolumn{3}{c|}{billion ILCU}  & billion JPY  & \multicolumn{2}{c|}{k FTE-years}\\
\hline
\end{tabular}
\end{center}
\label{cost-table}
\end{table}%

Estimates of the additional cost for three options have also been made. 
Increasing the centre-of-mass energy from 250 to \SI{500}{GeV} was estimated to be 3.9 to 4.2 billion ILCU for (Acc+CF) and 55 billion JPY for CE.
The electron-driven positron source, which might be needed as a backup plan for smooth ILC startup in case the undulator-driven positron source technology requires more time to mature, was estimated to be 0.20 billion ILCU for (Acc+CF) and 12.5 billion JPY for CE.
An additional interaction point was approximated to be on the order of 0.5 billion ILCU. This is the cost to add a second beam delivery system (BDS) and interaction point, excluding CE work which must be evaluated separately. In addition, new beam splitting systems need to be provided at both BDS upstream ends. Dedicated design work is needed to improve the accuracy of this approximate estimate.

\section{Other Considerations}
\subsection{Timescale}
Based on the development of the TDR and ongoing ITN effort, the ILC is technically mature and is ready for engineering design studies to prepare for project construction. Following completion of a four-year preparatory period, the construction period is estimated to be nine years, followed by one year of commissioning. Thus, the start of physics exploitation will be ten years after project start. 
\subsection{Environmental considerations}
The environmental impact of the ILC project has been extensively studied. In a combined effort, the ILC and CLIC projects commissioned two Lifecycle Assessment (LCA) studies with ARUP, a consultancy company with extensive experience in this area. The first study quantified the environmental impact of the civil construction of the underground tunnels, caverns, and shafts for the projects~\cite{ARUP-1}.
A second, more comprehensive study extended the assessment to include the environmental impact of the accelerator and detector materials and construction\cite{ARUP-2}. It was concluded
that 609 kilotons equivalent $\rm CO_2$ Green House Gas would be emitted by the construction. 

\subsection{Yearly power consumption}
An estimate of yearly power consumption of ILC operation was prepared. As indicated above, the total power needed during machine operation is estimated to be \SI{111}{MW}. It was assumed that there would be little difference in the power requirement between physics running and machine development periods, because machine development will constantly use the beams. It was assumed that the machine would be kept cold during the shutdown period in order to ensure machine stability and consume \SI{20}{MW}. With a yearly shutdown period of \SI{30}{\%}, this leads to a yearly power consumption of \SI{0.73}{TWh}. 

\section{Plan of ILC in Japan}
The ILC community effort in Japan has been driven by ILC-Japan\footnote{\url{https://ilc-japan.org/en/}}, overseen by JAHEP, and by KEK. The accelerator work and the physics and detector work are fully integrated in IDT Working Groups 2 and 3, respectively, where large contributions have been made by Japanese 
members. In accelerator development, KEK has been playing a leading role in the many work packages of the ITN. 

The Japanese HEP community has been addressing the concerns expressed by MEXT for pursuing the ILC as a global project hosted in Japan, which is explained in Section~\ref{section-2}. These efforts include developing synergies with other particle physics experiments in detector development, explaining the benefit for other sciences of technologies developed for the ILC, and 
maintaining 
continuous communication of scientific value to the wider community. 

Noting the different views
of the international community and MEXT
on how to proceed with the ILC, 
as identified by the IDT-IEP (see Section~\ref{section-2}),
ILC-Japan and KEK have adopted a phased approach. 
The first step is to establish an international agreement on the necessity of the ILC\footnote{It is noted that an electron-positron colliding beam Higgs factory, such as the ILC, has already been established as the highest priority future facility amongst the international high energy physics community. } through discussion among the potential partner countries. 
Topics of discussion would also include procedures for cost and responsibility sharing, organisation and governance structure, and site selection.
This step is expected to be decoupled from the choice of host nation and would take note of the status of other major proposals.

Once international agreement has been achieved, 
the project would be defined in the next phase by all interested countries following the procedure agreed in the first step. Discussion leading to adoption of the governance and cost-sharing model as well as the host country and site could then follow.
As part of establishing the project, a Pre-lab could be started. 
Together with other key players, ILC-Japan and KEK have been seeking 
a circumstance
where the government of Japan could play a leading role toward reaching an international consensus on the ILC.

This approach is in line with the view expressed by JAHEP in their document submitted for the European Strategy for Particle Physics 2026. The document 
states that the early realization of a Higgs factory through international collaboration is crucial for the future of the field and
\begin{quote}
`` {\itshape ..., we will pursue the following two key directions:}
\begin{itemize}
 \item {\itshape We prioritize efforts to realize the ILC as Global Project, taking a leading role in advancing ongoing
initiatives. We will engage with international partners to discuss governance, responsibilities, and site
selection. We intend to develop and expand our scientific and promotional activities to host the ILC as
Global Project in Japan}.
    \item  {\itshape We also extend our activities in other Higgs factory proposals as a collective approach to maximize the
chances of timely realizing a Higgs factory.}''
\end{itemize}
\end{quote}

\section{Summary}
The ILC is a candidate next-generation collider based on well-established SRF technology and a mature technical design and offering polarized beams. ILC will be a capable electron-positron Higgs factory in its first phase and could be upgraded to \SI{550}{GeV}
and 
further 
to 1 TeV using the same technology. Although the ILC has not yet gained political acceptance by the Japanese government, stewarded by the International Development Team and through the work of ILC-Japan and KEK, progress has been made since the Technical Design Report. This paper has briefly described 
that
progress. Planning has been completed for the next phase of development, the Preparatory Phase, during which technical activities and engineering design necessary for ILC construction will be completed. The framework, the ILC Technology Network hosted by KEK, and work plan have been established to advance development of key ILC technologies through international cooperation during the period preceding the Preparatory Phase. The International Expert Panel has completed a study of the challenges of initiating the ILC as a global project. Lifecycle assessments of the environmental impact of the ILC and estimate of the yearly power consumption have been completed. The estimate of the cost of the ILC as a Higgs factory operating at \SI{250}{GeV} has been updated to be 6.78 billion ILCU for accelerator and conventional facilities plus 196 billion JPY for civil engineering in Japan. The ILC accelerator design is site independent, and the ILC could be built at a site other than Japan. In the future, the infrastructure for the ILC could serve as the basis of a much higher energy and luminosity linear collider capitalising on advances in acceleration technologies.


\clearpage
\section*{The ICFA ILC International Development Team and ILC-Japan}
The IDT activities are being carried out by Working Groups for ``Accelerator'' and ``Detector and Physics''. ILC-Japan consists of working groups on accelerator R\&D, detector and physics, as well as a task force on public relations,  
{\footnotesize
\begin{multicols}{2}

\begin{center}{
R.~Dowd\\
}\textbf{Australian Synchrotron, Clayton, Victoria, Australia}
\end{center}
\vspace{-0.5cm}

\begin{center}{
G.~Taylor,
P.~Urquijo\\
}\textbf{University of Melbourne, Melbourne, Victoria, Australia}
\end{center}
\vspace{-0.5cm}

\begin{center}{
C.~Hensel\\
}\textbf{Centro Brasileiro de Pesquisas F\'isicas, Rio de Janeiro, Brazil}
\end{center}
\vspace{-0.5cm}

\begin{center}{
R.~Laxdal\\
}\textbf{TRIUMF, Wesbrook Mall, Vancouver, Canada}
\end{center}
\vspace{-0.5cm}

\begin{center}{
Y.~Gao\\
}\textbf{Peking University, Haidian District, Beijing, China}
\end{center}
\vspace{-0.5cm}

\begin{center}{
E.~Cenni,
C.~Madec,
O.~Napoly,
M.~Titov\\
}\textbf{Institut de recherche sur les lois fondamentales de l'Univers, CEA Saclay, Gif sur Yvette, France}
\end{center}
\vspace{-0.5cm}

\begin{center}{
A.~Faus Golfe,
R.~P\"oschl\\
}\textbf{IJCLab, Universit\'e Paris-Saclay, CNRS/IN2P3, Orsay, France}
\end{center}
\vspace{-0.5cm}

\begin{center}{
T.~Behnke,
K.~Buesser,
F.~Gaede,
K.~Kr\"uger,
J.~List,
B.~List,
S.~Riemann,
N.~Walker,
H.~Weise\\
}\textbf{Deutsches Elektronen-Synchrotron DESY, Hamburg, Germany}
\end{center}
\vspace{-0.5cm}

\begin{center}{
F.~Simon\\
}\textbf{Karlsruhe Institute of Technology, Karlsruhe, Germany}
\end{center}
\vspace{-0.5cm}

\begin{center}{
G.~Moortgat-Pick\\
}\textbf{University of Hamburg, Hamburg, Germany}
\end{center}
\vspace{-0.5cm}

\begin{center}{
K.~Mazumdar$^{0}$\\
}\textbf{Tata Institute of Fundamental Research, Navy Nagar, Colaba, Mumbai, Maharashtra, India}
\end{center}
\vspace{-0.5cm}

\begin{center}{
M.~Zobov\\
}\textbf{INFN Laboratori Nazionali di Frascati, Frascati, Italy}
\end{center}
\vspace{-0.5cm}

\begin{center}{
L.~Monaco\\
}\textbf{INFN Sezione di Milano, Milano, Italy}
\end{center}
\vspace{-0.5cm}

\begin{center}{
F.~Forti\\
}\textbf{INFN Sezione di Pisa, Pisa, Italy}
\end{center}
\vspace{-0.5cm}

\begin{center}{
T.~Kitahara\\
}\textbf{Chiba University, Inage-ku, Chiba, Japan}
\end{center}
\vspace{-0.5cm}

\begin{center}{
Y.~Abe,
S.~Arai,
H.~Araki,
S.~Araki,
Y.~Arimoto,
A.~Aryshev,
S.~Asai,
R.~Bajpai,
T.~Dohmae,
M.~Egi,
Y.~Enomoto,
K.~Fujii,
M.~Fukuda,
T.~Goto,
K.~Hanagaki,
K.~Hara,
T.~Hara,
M.~Hiraki,
T.~Honma,
H.~Ito,
D.~Jeans,
R.~Katayama,
S.~Kessoku,
T.~Koseki,
T.~Kubo,
K.~Kubo,
A.~Kumar,
M.~Kurata,
T.~Matsumoto,
S.~Michizono,
T.~Miura,
Y.~Morikawa,
H.~Nakai,
H.~Nakajima,
K.~Nakamura,
E.~Nakamura,
K.~Nakanishi,
M.~Nojiri,
H.~Oide,
Y.~Okada,
T.~Okugi,
M.~Omet,
K.~Popov,
T.~Saeki,
N.~Saito,
H.~Sakai,
M.~Sato,
S.~Shanab,
H.~Shimizu,
R.~Takahashi,
N.~Terunuma,
M.~Togawa,
R.~Ueki,
K.~Umemori,
Y.~Ushiroda,
E.~Viklund,
Y.~Watanabe,
T.~Yamada,
A.~Yamamoto,
Y.~Yamamoto,
M.~Yamauchi,
K.~Yokoya\\
}\textbf{KEK, Tsukuba, Ibaraki, Japan}
\end{center}
\vspace{-0.5cm}

\begin{center}{
M. ~Kuriki,
T.~Takahashi\\
}\textbf{Hiroshima University, Higashi-Hiroshima, Hiroshima, Japan}
\end{center}
\vspace{-0.5cm}

\begin{center}{
O.~Jinnouchi\\
}\textbf{Institute of Science Tokyo, Meguro-ku, Tokyo, Japan}
\end{center}
\vspace{-0.5cm}

\begin{center}{
S.~Narita\\
}\textbf{Iwate University, Morioka, Iwate, Japan}
\end{center}
\vspace{-0.5cm}

\begin{center}{
J.~Maeda\\
}\textbf{Kobe University, Nada-ku, Kobe, Japan}
\end{center}
\vspace{-0.5cm}

\begin{center}{
S.~Chen,
T.~Nakaya\\
}\textbf{Kyoto University, Sakyo-ku, Kyoto, Japan}
\end{center}
\vspace{-0.5cm}

\begin{center}{
K.~Kawagoe,
K.~Tsumura\\
}\textbf{Kyushu University, Nishi-ku, Fukuoka, Japan}
\end{center}
\vspace{-0.5cm}

\begin{center}{
Y.~Horii,
K.~Inami\\
}\textbf{Nagoya University, Chikusa-ku, Nagoya, Aichi, Japan}
\end{center}
\vspace{-0.5cm}

\begin{center}{
M.~Iwasaki\\
}\textbf{Osaka Metropolitan University, Sumiyoshi-ku, Osaka, Japan}
\end{center}
\vspace{-0.5cm}

\begin{center}{
T.~Masubuchi\\
}\textbf{Osaka University, Toyonaka, Osaka, Japan}
\end{center}
\vspace{-0.5cm}

\begin{center}{
T.~Sanuki\\
}\textbf{The Jikei University School of Medicine, Chofu, Tokyo, Japan}
\end{center}
\vspace{-0.5cm}

\begin{center}{
M.~Ishino,
T.~Mori,
W.~Ootani,
T.~Suehara,
J.~Tian$^{1}$\\
}\textbf{The University of Tokyo, Bunkyo-ku, Tokyo, Japan}
\end{center}
\vspace{-0.5cm}

\begin{center}{
S.~Hirose\\
}\textbf{University of Tsukuba, Tsukuba, Ibaraki, Japan}
\end{center}
\vspace{-0.5cm}

\begin{center}{
A.F.~\.Zarnecki\\
}\textbf{Faculty of Physics, University of Warsaw, Warsaw, Poland}
\end{center}
\vspace{-0.5cm}

\begin{center}{
I.~Bozovic\\
}\textbf{Vin\v{c}a Institute of Nuclear Sciences, University of Belgrade, Belgrade, Serbia}
\end{center}
\vspace{-0.5cm}

\begin{center}{
L.~Garc\'ia-Tabar\'es,
C.~Oliver Amor\'os\\
}\textbf{Centro de Investigaciones  Energ\'eticas, Medioambientales y Tecnol\'ogicas (CIEMAT), Madrid, Spain}
\end{center}
\vspace{-0.5cm}

\begin{center}{
M.~Vos\\
}\textbf{IFIC, CSIC-Universitat de Val\`encia, Valencia, Spain}
\end{center}
\vspace{-0.5cm}

\begin{center}{
N.~Catalan-Lasher\'as,
D.~Delikaris,
S.~Doebert,
A.~Latina,
J.~Osborne,
P.~Sievers,
S.~Stapnes\\
}\textbf{CERN, Geneva, Switzerland}
\end{center}
\vspace{-0.5cm}

\begin{center}{
T.~Nakada\\
}\textbf{High Energy Physics Laboratory, Institute of Physics, Ecole Polytechnique F\'ed\'erale de Lausanne, Lausanne, Switzerland}
\end{center}
\vspace{-0.5cm}

\begin{center}{
J.~Clark,
P.A.~McIntosh\\
}\textbf{STFC Daresbury Laboratory, Warrington, United Kingdom}
\end{center}
\vspace{-0.5cm}

\begin{center}{
A.~Robson\\
}\textbf{University of Glasgow, Glasgow, United Kingdom}
\end{center}
\vspace{-0.5cm}

\begin{center}{
P.N.~Burrows\\
}\textbf{University of Oxford, Oxford, United Kingdom}
\end{center}
\vspace{-0.5cm}

\begin{center}{
J.~Zhang\\
}\textbf{Argonne National Laboratory, Lemont, IL, USA}
\end{center}
\vspace{-0.5cm}

\begin{center}{
D.~Denisov,
B.~Parker\\
}\textbf{Brookhaven National Laboratory, Upton, NY, USA}
\end{center}
\vspace{-0.5cm}

\begin{center}{
G.~Dugan,
M.~Liepe,
J.R.~Patterson,
D. L.~Rubin\\
}\textbf{Cornell University, Ithaca, NY, USA}
\end{center}
\vspace{-0.5cm}

\begin{center}{
S.~Belomestnykh$^{2}$,
P.~McBride,
S.~Posen,
N.~Solyak\\
}\textbf{Fermi National Accelerator Laboratory, Batavia, IL, USA}
\end{center}
\vspace{-0.5cm}

\begin{center}{
R.L.~Geng,
J.~Grames,
R.A.~Rimmer,
R.~Ruber\\
}\textbf{Jefferson Lab, Newport News, VA, USA}
\end{center}
\vspace{-0.5cm}

\begin{center}{
M.~Demarteau\\
}\textbf{Oak Ridge National Laboratory, Oak Ridge, TN, USA}
\end{center}
\vspace{-0.5cm}

\begin{center}{
T.W.~Markiewicz,
M.E.~Peskin,
M. C.~Ross,
C.~Vernieri\\
}\textbf{SLAC National Accelerator Laboratory, Menlo Park, CA, USA}
\end{center}
\vspace{-0.5cm}

\begin{center}{
H.~Murayama$^{3}$\\
}\textbf{University of California, Berkeley, CA, USA}
\end{center}
\vspace{-0.5cm}

\begin{center}{
A. J.~Lankford\\
}\textbf{University of California, Irvine, CA, USA}
\end{center}
\vspace{-0.5cm}

\begin{center}{
J.E.~Brau,
J.~Strube\\
}\textbf{University of Oregon, Eugene, OR, USA}
\end{center}
\vspace{-0.5cm}

\begin{center}{
A.P.~White\\
}\textbf{University of Texas at Arlington, Arlington, TX, USA}
\end{center}
\vspace{-0.5cm}

\begin{flushleft}
{$^{0}$}Retired on 31 October 2024.\\
{$^{1}$}Also at: ICEPP\\
{$^{2}$}Also at: Physics and Astronomy, Stony Brook University, NY 11794, USA\\
{$^{3}$}Also at:Kavli Institute for the Physics and Mathematics of the Universe, University of Tokyo
\end{flushleft}

\end{multicols}
}

\end{document}